\documentclass[12pt]{iopart}
\usepackage{iopams}  

\usepackage{graphicx}
\usepackage{subcaption}
\usepackage{xcolor}
\usepackage{ulem}

\begin{document}


\title[Vectorial probing of optical transitions]{Vectorial probing of electric and magnetic transitions in variable optical environments and {\it vice-versa}}

\author{Reinaldo Chacon$^1$, Aymeric Leray$^1$, Jeongmo Kim$^2$, Khalid Lahlil$^2$, Alexandre Bouhelier$^1$, Jong-Wook Kim$^2$, Thierry Gacoin$^2$, G\'erard {Colas des Francs}$^1$}
\address{$^1$Laboratoire Interdisciplinaire Carnot de Bourgogne (ICB), UMR 6303 CNRS, Universit\'e Bourgogne Franche-Comt\'e, 9 Avenue Savary, BP 47870, 21078 Dijon Cedex, France}
\address{$^2$Physique de la Mati\`ere Condens\'ee, CNRS UMR 7643, Ecole Polytechnique, 91128 Palaiseau, France
}
\ead{gcolas@u-bourgogne.fr}

\begin{abstract}
We use europium doped single crystalline NaYF$_4$ nanorods for probing the electric and magnetic contributions to the local density of optical states (LDOS). Reciprocically, we determine intrinsic properties of the emitters (oscillator strength, quantum yield) by comparing their measured and simulated optical responses in front of a mirror. We first experimentally determine the specifications of the nanoprobe (orientation and oscillator strength of the electric and magnetic dipoles moments) and show significant orientation sensitivity of the branching ratios associated with electric and magnetic transitions. In a second part, we measure the modification of the LDOS in front of a gold mirror in a Drexhage's experiment. We discuss the role of the electric and magnetic LDOS on the basis of numerical simulations, taking into account the orientation of the dipolar emitters. We demonstrate that they behave like degenerated dipoles sensitive to polarized partial LDOS.
\end{abstract}

%
\vspace{2pc}
\noindent{\it Keywords}: Magnetic dipole, LDOS, nanophotonics, optical near-field probe\\
%
\submitto{\NT}
%
%
%

\section{Introduction}
The electromagnetic local density of states (LDOS) is a key parameter governing light-matter interaction processes such as radiative heat transfer \cite{joulain05,Moskalensky-Yurkin:21} or spontaneous emission \cite{Barnes1998b,GCF-Barthes-Girard:2016,Wojszvzyk:19} so that it is of importance for controlling heat dissipation in optoelectronics devices \cite{Cuevas-GarciaVidal:18}, or coherence emission of single photon source for quantum technologies \cite{Lodahl:2015}. Therefore, strong efforts have been put forward to map and control LDOS in the optical near-field \cite{Dereux-Devaux-Weeber-Goudonnet-Girard:2001,NanotechDereuxGCF:2003,deWilde:06,HoogenboomNanoLet:2009,Agio:2012,Kasperczyk:15,Pham-Drezet:16} with more particular attention devoted to the role of electric and magnetic contributions in the last decade \cite{Karaveli-Zia:2011,Aigouy:14,Rabouw-Norris:16,Li-Zia:18,Wiecha-Cuche:19,Bidault-Mivelle-Bonod:19}. Indeed, although light-matter interaction is mainly governed by electric dipole in the visible range, it is possible to enhance magnetic dipole contribution, opening a new parameter to control light at the nanoscale.These pionered works considered randomly oriented rare-earth ions as local probes of the electromagnetic environment. They mainly investigated fluorescence intensity, giving access to the so-called {\it radiative} LDOS, either electric or magnetic \cite{Karaveli-Zia:2011,Aigouy:14,Rabouw-Norris:16,NorrisPRL:18}. The full LDOS, including radiative and non radiative channels for both electric and magnetic contributions, is proportional to fluorescence decay rate so that it can be probed by lifetime measurement.  In 2018, Zia and coworkers measured the lifetime of ions in front of a gold mirror or dielectric interfaces and discussed the  electric and magnetic contribution to the complete LDOS \cite{Li-Zia:18}. They used randomly oriented emitters doping thin films, probing the LDOS averaged over the field polarizations.

In this work, we consider individual europium doped NaYF$_4$ (NaYF$_4$:Eu$^{3+}$) nanorods as quasi vectorial probes of the electric and magnetic LDOS. Eu$^{3+}$ ions present both electric and magnetic dipole transitions in the optical range so that the lifetime of their excited states are sensitive to both electric and magnetic LDOS. We recently characterized the electric and magnetic optical transitions in single crystalline NaYF$_4$:Eu$^{3+}$ nanorods  and demonstrated well defined orientation of the associated dipole moments \cite{Chacon-GCF:2020,Kim-Gacoin:21}. Moreover, the  low phonon ($300-400$ cm$^{-1}$) NaYF$_4$ host matrix ensures narrow electric and magnetic emission lines and facilitates their distinction in the spectra \cite{Rabouw-Norris:16,Suyver-Gudel:06,Chacon-GCF:2020,Kim-Gacoin:21,Kumar-Fick:20}.
Finally, single crystalline host matrix induces well defined polarized electric dipole (ED) or magnetic dipole (MD) optical transitions, related to the nanocrystal c-axis orientation \cite{Sayre-Freed:1956,Kim-Gacoin:17}. The dipole transition presents a fixed orientation with respect to the rod axis and constitutes a degenerated dipole of interest for {\it vectorial} nanoprobing.  Reversely, analyzing the modification of the fluorescence signal in complex environment enables to access the intrinsic properties of the emitters such as orientation \cite{Sick-Hecht-Wild-Novotny:2001}, quantum yield \cite{Buchler-Kalkbrenner-Hettich-Sandoghdar:2005} and oscillator strength \cite{Aigouy:14}.

In this article, we take benefit of these engineered optical nanosources to investigate electric and magnetic LDOS variations. In section \ref{sect:PolaDiag}, we determine the ED/MD orientation from polarized diagram emission measurement. Then, in section \ref{sect:Branching}, we quantify oscillator strengths of the emitters by measuring the ED/MD branching ratio as a function of the distance to a gold mirror. Having fully characterized intrinsic optical properties of NaYF$_4$:Eu$^{3+}$ nanocrystal, we use it as a LDOS nanoprobe in a Drexhage-like experiment 
in section \ref{sect:LDOS}. 

\section{Specifications of the optical properties of NaYF$_4$:Eu$^{3+}$ nanoprobe}  
\subsection{Electric and magnetic dipole moments orientation}
\label{sect:PolaDiag}

\begin{figure*}[h!]
\centering
\includegraphics[width=10cm]{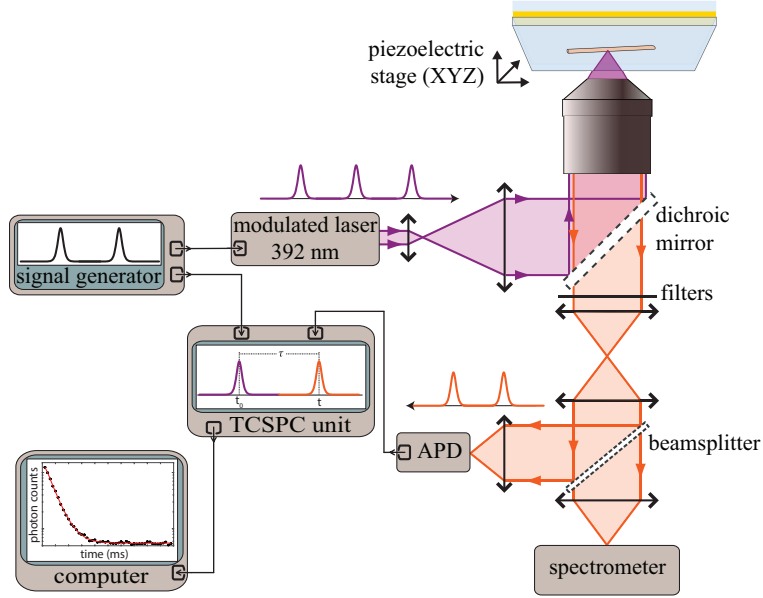}
	\caption{Schematic representation of the confocal optical microscope. The europium ions are excited by a laser emitting at a wavelength of 392 nm CW or modulated at  10 Hz) and focused by a microscope objective (40×, NA = 0.6, Nikon). The luminescence is collected by the same objective and is separated from the excitation laser by a dichroic mirror and routed either to an avalanche photodiode (APD) for confocal imaging or to a spectrometer. The APD is connected to a time correlated single photon counting card (TCSPC; Picoquant-PicoHarp 300) for fluorescence decay acquisition.
   		\label{fig:lifetimeSetup}}
\end{figure*}

\begin{figure*}[h!]
\centering
	\includegraphics[width=15cm]{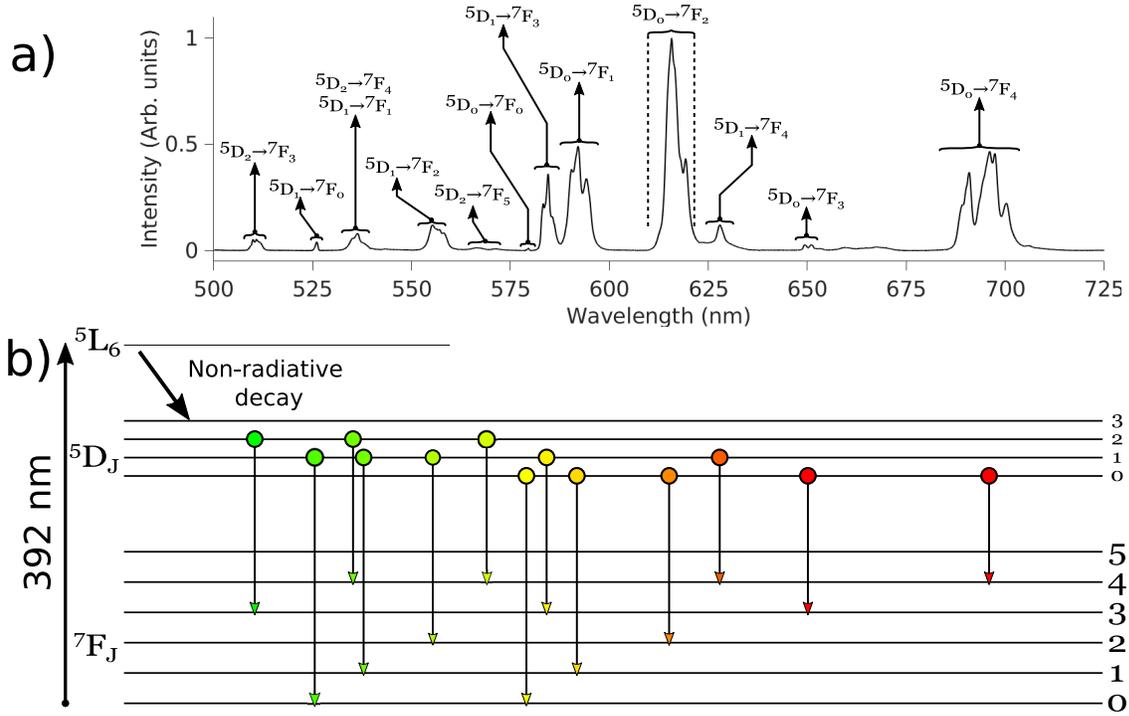}
	\caption{a) Eu$^{3+}$ doped NaYF$_4$ nanorods emission spectrum taken without analyzer. b) Pictorial representation of the electronic transitions corresponding to the observed spectral lines. All the optical transitions are of ED nature except $^5D_0\rightarrow ^7F_ 1$, 
	$^5D_1\rightarrow ^7F_ 2$ and $^5D_2\rightarrow ^7F_ 3$ that are of MD nature \cite{Rabouw-Norris:16,Binnemans:96}.
		\label{fig:FullSpec}}
\end{figure*}

\begin{figure*}
	\includegraphics[width=17cm]{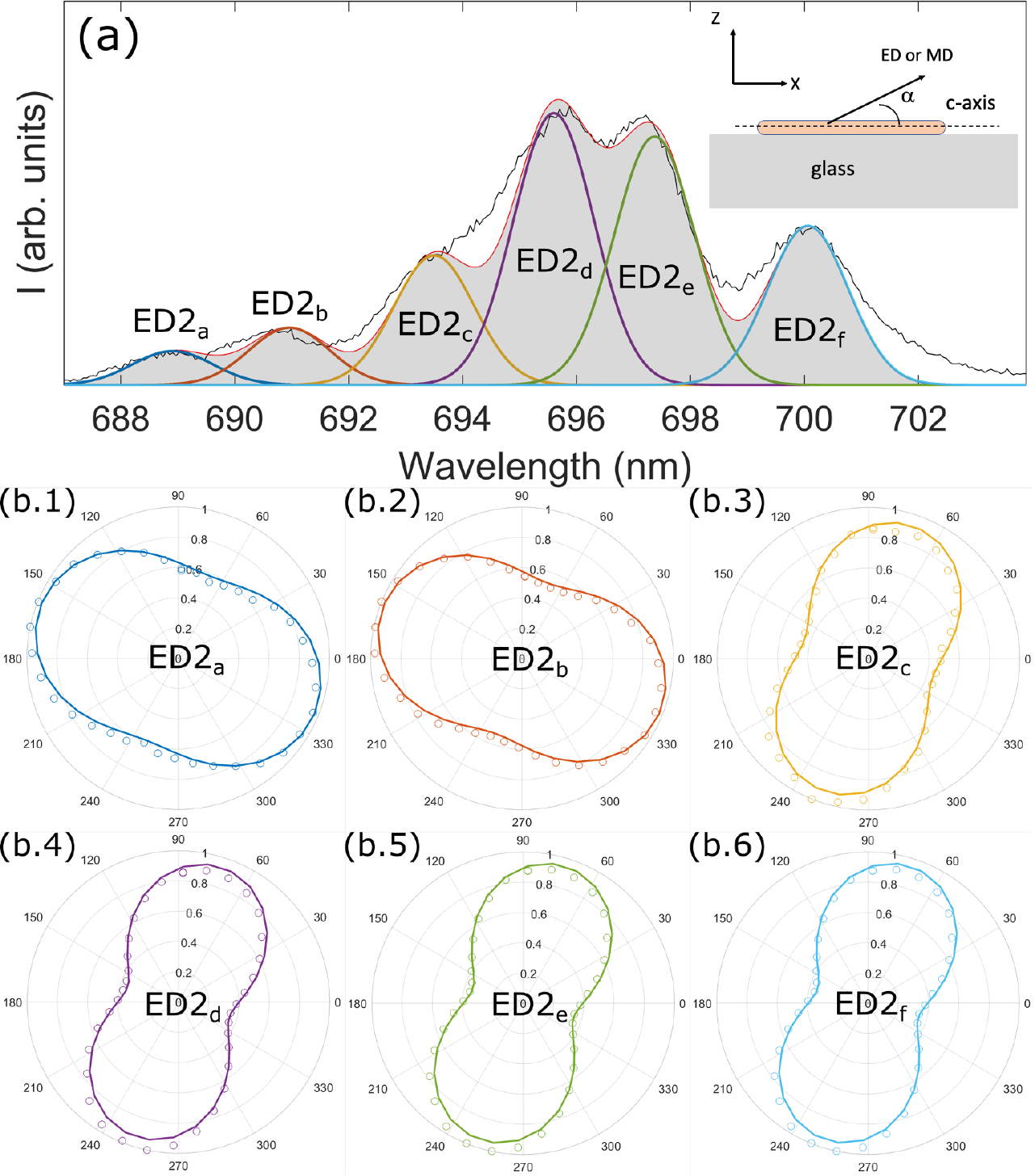}
	\caption{(a) Luminescence spectrum of single NaYF$_4$:Eu$^{3+}$ nanorod in the spectral range corresponding to $^5$D$_0\rightarrow ^7$F$_4$ transition. Spectral unmixing  is performed by fitting experimental data with a sum of six Gaussians.  The configuration is schemed in the inset. (b1) to (b5) The normalized emission  intensity diagrams of the six  ED optical transitions are shown with dots, as a function of the analyser orientation. For each dipole, normalization is performed on the maximum intensity of single nanorod. The lines correspond to the theoretical fits using the full model of ref. \cite{Chacon-GCF:2020}.
\label{fig:PolaDiagF4}}
\end{figure*}

Single crystal NaYF$_4$:Eu$^{3+}$ nanorods, with average length and diameter of $1.1 \mu m$ and 120 nm respectively, were synthesized using hydrothermal procedure described in references \cite{Wang-Liu:2010,Chacon-GCF:2020}. The doped nanorods were deposited by drop-casting on a quartz coverslip bearing a grid landmark pattern. The emission spectra of single nanorod were acquired with a home-made confocal microscope, schemed in Fig. \ref{fig:lifetimeSetup}. The europium ions were excited by a continuous-wave (CW) laser at a wavelength of $392\pm 0.5$ nm, focused by a low numerical aperture objective lens (NA=0.60, 40X, Nikon). The photoluminescence emission was collected through the same objective. After collection, the light was directed towards an avalanche photodiode (APD) or a spectrometer. Single nanorod spectrum is presented in Fig. \ref{fig:FullSpec} with the corresponding transitions lines. In the following, we focus on the three emission lines originating from the excited state $^5$D$_0$: $^5D_0\rightarrow ^7F_ 1$, $^5D_0\rightarrow ^7F_ 2$ and $^5D_0\rightarrow ^7F_ 4$. Indeed, we need to characterize the dipole moments of all the optical transitions involved in the relaxation of the excited state $^5D_0$ before using it as LDOS nanoprobe. This notably requests to know the nature (electric or magnetic), the orientation and the numbers (associated to the $^7$F$_J$ levels and their Stark sublevels) of the dipole moments. These properties strongly depends on the host matrix. The MD ($^5D_0\rightarrow ^7F_ 1$) and ED ($^5D_0\rightarrow ^7F_ 2$) optical transitions have been extensively investigated in ref. \cite{Chacon-GCF:2020,Kim-Gacoin:21}. Therefore, we summarize our method to characterize the dipolar transition moment considering the ED ($^5D_0\rightarrow ^7F_ 4$) transition only, for the sake of clarity. Dipole moment orientation can be deduced from polarized emission diagram. We recorded polarized emission spectra of individual rods, placing an analyser in front of the spectrometer, see Fig. \ref{fig:PolaDiagF4}. Scanning electron microscopy (SEM) was systematically performed after optical characterizations and only data acquired for individual rods are kept in the analysis. The crystal-field perturbation lift the degeneracy of the $^7F_J$ levels into to (2J+1) Stark sublevels \cite{Tu-Chen:13}. In Fig. \ref{fig:PolaDiagF4}a) we observe six of the nine ED transitions  ($^5D_0\rightarrow ^7F_ 4$), noted ED$_{a, \ldots , f}$. The spectra are fitted with a sum of six Gaussian functions and the contribution of each peak (area under each Gaussian) is plotted as a function of the angle of the analyser in Fig.  \ref{fig:PolaDiagF4}b).

Optical emission is modeled considering a single rod with its c-axis aligned with the substrate/air interface. The ED or MD moments associated to a transition line present a fixed angle $\alpha$ with respect to the c-axis inducing that the rod emission is equivalent to the incoherent emission of three electric or magnetic orthogonal dipoles \cite{Chacon-GCF:2020}
\begin{eqnarray}
{\bf p}_1&=&
\left(
\begin{tabular}{c}
$\cos\alpha$ \\ $0$ \\ $0$
\end{tabular}
\right) \,,
{\bf p}_2=\frac{1}{\sqrt{2}}
\left(
\begin{tabular}{c}
$0$ \\ $\sin \alpha$  \\$0$ 
\end{tabular}
\right)\,,
{\bf p}_3=\frac{1}{\sqrt{2}}
\left(
\begin{tabular}{c}
$0$ \\$0$ \\ $\sin\alpha$
\end{tabular}
\right) \,.
\end{eqnarray}
The dipolar angles $\alpha$ are determined from polarized diagram emission \cite{Kim-Gacoin:17,Chacon-GCF:2020,Kim-Gacoin:21}. For low NA objective, the detected intensity  is expressed as  
\begin{eqnarray}
I(\Phi_A)=A \sin^2\Phi_A +B \cos^2\Phi_A
\label{eq:paraxial}
\end{eqnarray}
where $\Phi_A$ refers to the angle of the analyzer with respect to the rod c-axis. $A=\sin^2 \alpha/2, B=\cos^2 \alpha$ for an electric dipole and  $A=\cos^2 \alpha, B=\sin^2 \alpha/2$ for a magnetic dipole. So, the emitter angle with respect to the c-axis can be estimated from the fitting parameters $A,B$ of the measured polarized diagram. We have used exact expressions considering the NA of the objective for the sake of accuracy but the paraxial approximation can be safely used for $NA<0.7$. The fits are presented as solid lines in Fig.\ref{fig:PolaDiagF4}b), with the parameters gathered in Table \ref{angles}. ED2$_a$ and ED2$_b$ present similar angles with respect to the c-axis and close emission wavelengths ($\lambda \simeq 690$ nm, $ \alpha \simeq 63 ^\circ$) so that they can be treated as a unique electric  dipole (noted ED2$_{ab}$) in the following. Dipole moments ED2$_c$ to ED2$_f$ are all emitting around 697 nm and are oriented close to $\alpha \simeq 40 ^\circ$. Note that it can be sometimes difficult to unambiguously determine the angle $\alpha$ leaving both c-axis and $\alpha$ orientations as free parameters. However, we carefully determined the nanorod orientation and MD angle by confocal microscopy and took care that c-axis orientation is the same for ED and MD transitions, lifting the ambiguity.  See ref. \cite{Chacon-GCF:2020} for details.  

We applied the same procedure for ED and MD transitions for several Eu$^{3+}$doped nanorods. The results are summarized in Table \ref{angles}.  We observed three MD transitions ($^5D_0\rightarrow ^7F_ 1$) as expected because of the crystal field splitting of the Stark sub-levels. MD1 ($\lambda=590$ nm, $\alpha=68 \pm 1^\circ$) and MD2 ($\lambda=592$ nm, $\alpha=72 \pm 1^\circ$) present similar emission properties so that they can be treated as a unique magnetic dipole (MD12). MD3 ($\lambda=595$ nm, $\alpha=37.3 \pm 0.5^\circ$) has to be considered separately. We also resolved two of the five ED $^5D_0\rightarrow ^7F_ 2$ transitions. Finally, as pointed above, we observed six of the nine  ED transitions  ($^5D_0\rightarrow ^7F_ 4$) but again with only two different angles.

\begin{table} [!ht]
\caption{\label{angles}Dipole moment angle of transitions $^5$D$_0\rightarrow ^7$F$_1$, $^5$D$_0\rightarrow ^7$F$_2$ and $^5$D$_0\rightarrow ^7$F$_4$. $\alpha$ defines the angle of oscillation with respect to the c-axis of the electric and magnetic dipoles deduced from spectral analysis of single NaYF$_4$:Eu$^{3+}$ nanorods. Similar dipoles present the same emission properties and can be gathered in a simplified model (line "approx").}

\begin{tabular}{c|ccc|cc}
transition & \multicolumn{3}{c}{$^5D_0\rightarrow$ $^7F_ 1$} & \multicolumn{2}{c}{$^5D_0\rightarrow ^7F_ 2$} \\
 \br
dipole &MD1 & MD2 	& MD3  & ED1a 	& ED1b \\ 
&(590 nm)  & (592 nm) 	& (595 nm) & (615 nm) 	& (620 nm)	\\ 
\br
$\alpha$ &68$\pm 1^\circ$ &	72$\pm 1^\circ$					&37.3$\pm 0.5^\circ$ 						& 60.3 $\pm0.4^\circ$	&44.3$\pm 1.6^\circ$ \\
 & \multicolumn{2}{c}{\upbracefill} & & &  \\ 
 approx & \multicolumn{2}{c}{{MD}$_{12}$ (590 nm)}  & MD$_3$ (595 nm) & & \\
& \multicolumn{2}{c}{70$^\circ$}  &37$^\circ$& & 
\\
\end{tabular}
%
\begin{tabular}{c|cccccc}
\\
transition & \multicolumn{6}{c} {$^5D_0\rightarrow ^7F_ 4$} \\
\br
dipole  & ED2a & ED2b	& ED2c 	& ED2d	& ED2e & ED2f\\ 
 & (689 nm) 	& (691 nm)& (693 nm) & (696 nm) 	& (697 nm)& (700 nm) 		\\ 
\br
$\alpha$ & 63.4 $\pm 0.4^\circ$	&62.9$\pm 0.1^\circ$ & 43.3$\pm 0.5^\circ$ &40.7$\pm 0.3^\circ$&38.7$\pm 0.1^\circ$ & 38.8$\pm 0.2^\circ$\\
  &\multicolumn{2}{c}{\upbracefill} &\multicolumn{4}{c}{\upbracefill}\\ 
 approx & \multicolumn{2}{c}{{ED2}$_{ab}$ (690 nm)} &\multicolumn{4}{c}{{ED2}$_{cdef}$ (697 nm)}\\
  & \multicolumn{2}{c}{63$^\circ$}  &  \multicolumn{4}{c}{40$^\circ$}
\end{tabular}
\label{tab:Table1}
\end{table}

\subsection{Oscillator strengths of the ED and MD transitions}
\label{sect:Branching}
\begin{figure*}
	\includegraphics[width=17cm]{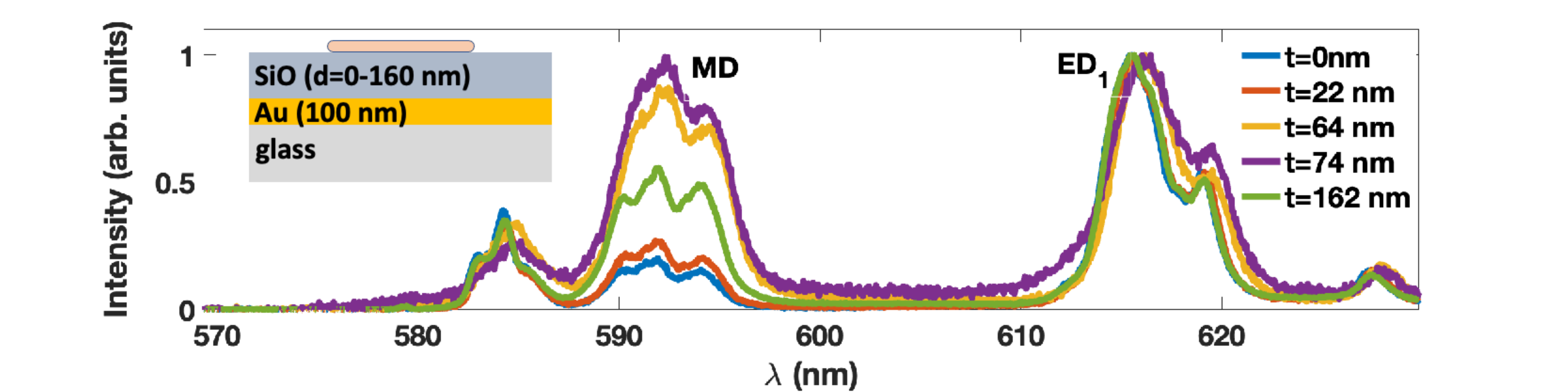}
	\caption{Luminescence spectra of single NaYF$_4$:Eu$^{3+}$ nanorod obtained at different distances to the gold mirror. The configuration is schemed in the inset. 
		\label{fig:RodSpectra}}
\end{figure*}

Having fully determined the orientation of the ED/MD lines, we now focus on measuring the ED/MD branching ratios for nanorods deposited on a gold mirror.  This gives access to the oscillator strengths of the optical transitions, of importance to quantify the LDOS probing. A grid landmark pattern was microfabricated on glass coverslips. Over this pattern, a layer of gold was deposited by thermal evaporation to create optically thick mirrors (thickness greater than 100 nm). On the gold mirrors a film of SiO was deposited by thermal evaporation to act as a spacer. The SiO spacer thickness was measured with a stylus profilometer (Dektak 150). The samples were cleaned with oxygen plasma and afterwards, a dispersion of rods in water was drop casted on the SiO film and left to dry. Fig. \ref{fig:RodSpectra} presents the spectra measured at different distances to the mirror. We observe a strong dependency of the MD signal with respect to the distance to the mirror, indicating a large modification of the MD/ED branching ratio.

In the previous section, we noticed that ED$_1$ and ED$_2$ transitions share very similar dipole orientation. Moreover, the optical signal poorly depends on the wavelength in the very near-field of the mirror were quasi-static dipole image reproduces the main features. So we simplify the discussion merging ED$_1$ and ED$_2$ contributions, see table \ref{tab:Table2}.

\begin{table} [!ht]
\caption{\label{angles2} Simplified model for ED/MD in NaYF$_4$:Eu$^{3+}$ nanorods.}
\begin{tabular}{c|cc|c|c}
 transition &MD$_{12}$ (590 nm)  & MD$_3$ (595 nm) & ED$_a$ & ED$_b$ \\
 \br 
$\alpha$ & 70$^\circ$  & 37$^\circ$& 63$^\circ$  & 40$^\circ$ \\
\end{tabular}
\label{tab:Table2}
\end{table}

We measure the emission spectra without analyser. Spectra are again fitted by gaussian profiles and the weight of each transition (MD12, MD3, EDa, EDb) is the surface area below the corresponding gaussian (MD12 weight is the sum of MD1 and MD2 peaks areas). Finally, each transition branching ratio is estimated. For instance
\begin{eqnarray}
\beta_{MD12}^{exp}=\frac{a_{MD12}}{a_{MD12}+a_{MD3}+a_{EDa}+a_{EDb}}
\end{eqnarray} 
where $a_{MD12}$ is the area of MD12 transition. Similar definitions occur for the other transitions. 

\begin{figure*}[h!]
\centering
	\includegraphics[width=8cm]{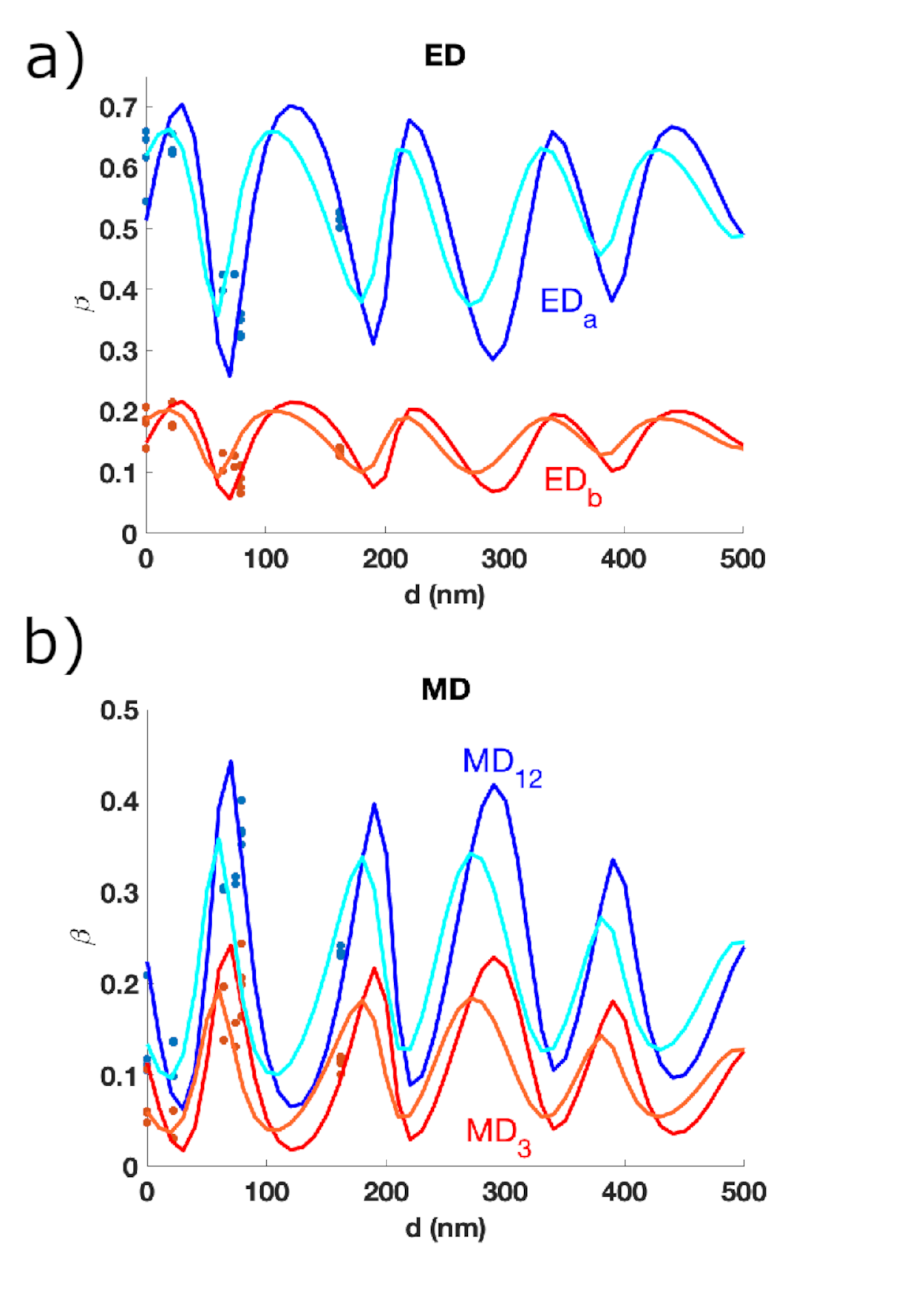}
	\caption{Branching ratios as a function of the distance to the mirror for (a) ED and (b) MD transitions. Dots correspond to experimental datas, solid lines are the fits taking into account size dispersion of the rod diameter between 90 and 130 nm light and dark colors correspond to the extremal rod diameters). Optical indices of SiO and Au are taken from ref. \cite{Hass-Salzberg:54,Johnson-Christy:1972}, respectively.  
		\label{fig:BranchingEDMD}}
\end{figure*}

The intensity recorded in the microscope objective is modelized as the incoherent emission of two dipoles parallel to the interface (${\bf p}_1$ and ${\bf p}_2$) and one dipole perpendicular to the interface (${\bf p}_3$). Therefore, radiative intensity at angle $\theta$ for an electric dipole transition is equal to 

\begin{eqnarray}
\label{eq:IED}
I^{ED}(\theta)&=&(p_1^2+p_2^2) I_{//}^{ED}(\theta)+p_3^2 I_\perp^{ED}(\theta) \\
\nonumber
&=&\left(\cos^2\alpha+\frac{1}{2}\sin^2\alpha \right) I_{//}^{ED}(\theta)+\frac{1}{2}\sin^2\alpha ~ I_\perp^{ED}(\theta) \\
\nonumber
I_{//}^{ED}(\theta)&=&\frac{3}{8}\left(\vert r_p\vert^2\cos^2\theta+\vert r_s\vert^2\right)+\frac{3}{4}Re\left[(-r_p \cos^2\theta+r_s)e^{2ik_0\cos \theta z_0}\right]\\
\nonumber
I_\perp^{ED}(\theta)&=&\frac{3}{4} \left[ \vert r_p\vert^2+2Re\left(r_pe^{2ik_0\cos \theta z_0}\right)\right]\sin^2\theta
\nonumber
\end{eqnarray}

$I_{//}$ and $I_\perp$ refer to the emission for a dipole parallel or perpendicular to the surface, respectively \cite{Lukosz1979,GCF-Barthes-Girard:2016}. $r_s$ and $r_p$ are the $s,p$-polarized Fresnel reflexion coefficients on the air/SiO(thickness $d$)/Au (100 nm)/glass multilayer substrate. $z_0$ refers to the altitude of the emitter. 

The magnetic dipole intensity is straightforwardly obtained with the exchange $r_p \leftrightarrow  r_s$: 

\begin{eqnarray}
\label{eq:IMD}
I^{MD}(\theta)&=&\left(\cos^2\alpha+\frac{1}{2}\sin^2\alpha \right) I_{//}^{MD}(\theta)+\frac{1}{2}\sin^2\alpha ~ I_\perp^{MD}(\theta) \\
\nonumber
I_{//}^{MD}(\theta)&=&\frac{3}{8}\left(\vert r_s\vert^2\cos^2\theta+\vert r_p\vert^2\right)+\frac{3}{4}Re\left[(-r_s \cos^2\theta+r_p)e^{2ik_0\cos \theta z_0}\right]\\
\nonumber
I_\perp^{MD}(\theta)&=&\frac{3}{4} \left[ \vert r_s\vert^2+2Re\left(r_s e^{2ik_0\cos \theta z_0}\right)\right]\sin^2\theta
\end{eqnarray}

The light collected throughout an objective of numerical aperture NA=0.6 (corresponding to an acceptance angle $\theta_{NA}=\sin^{-1}(NA/n_1)=37^\circ$ ), for a spacer thickness $d$ is 
\begin{eqnarray*}
I(d)=\int_{z_0=d}^{d+h}dz_0\int_{\theta=0}^{\theta_{NA}}I(\theta) \sin\theta d\theta 
\end{eqnarray*}
where we integrate the radiative intensity over the NA and the rod diameter $h$. Branching ratios can be numerically fitted from this model. For instance, 
\begin{eqnarray}
\nonumber \\
\beta_{MD12}(d)=\frac{f_{MD12}I_{MD12}(d) }{ f_{MD12}I_{MD12}(d)+f_{MD3}I_{MD3}(d)+f_{a}I_{EDa}(d)+f_{b}I_{EDb}(d)}
\end{eqnarray}  
$f_i$ is the oscillator strength of the transition $i$. Because ratio are fitted, we have only access to relative quantities $f_i$.  Fits are presented in Figure \ref{fig:BranchingEDMD} and fitting parameters are given in table \ref{OscStrength}. 

\begin{table} [!ht]
\caption{\label{OscStrength}Relative oscillator strengths obtained from the fitting of the branching ratios}
\begin{tabular}{c|cccc}
transition &MD12 	& MD3  & EDa 	& EDb		\\ 
\\ 
\hline
$f_i$ & 0.34	& 0.13	& 0.77	&0.19	\\
\end{tabular}
\label{tab:Table1}
\end{table}
We observe a fair agreement between the experimental data and our simulations, confirming the strong dependency of the electric and magnetic branching ratios with respect to the distance to the metallic mirror \cite{Karaveli-Zia:2011,Aigouy:14,Rabouw-Norris:16,Li-Zia:18}. The spatial oscillations  ($\sim 100$ nm) originate from interferences between direct dipolar emission and light reflected on the mirror (term $e^{2ik_0\cos \theta z_0}$ in Eqs. \ref{eq:IED},\ref{eq:IMD}).  Moreover, we clearly observe strong differences depending on the dipole transition orientation. Namely, the branching ratio of the electric transition oscillates between 30\% and 70 \% for  ED$_a$ ($\alpha=63^\circ$) and   between 5\% and 25 \% for  ED$_b$ ($\alpha=40^\circ$). Magnetic branching ratios vary in the range 5\% -45\% for MD$_{12}$ ($\alpha=70^\circ$) and 2-25 \% for  MD3 ($\alpha=37^\circ$), opening the door towards {\it vectorial} probing of the electromagnetic properties. Different branching ratios for ED and MD originate from different coupling efficiency to the surface plasmon polariton depending on the dipole nature, orientation and altitude. It is worth noticing that the orientation of the dipolar emitters is defined by the angle $\alpha$ with respect to the c-axis. Engineering the matrix host shape and anisotropy would lead  to different ED or MD polarizations for vectorial probing.

\section{Nanoprobing of the electric and magnetic local density of states}
\label{sect:LDOS}
\subsection{Relaxation channels for electric and magnetic dipolar emissions}
\begin{figure*}[h!]
\centering
\includegraphics[width=8cm]{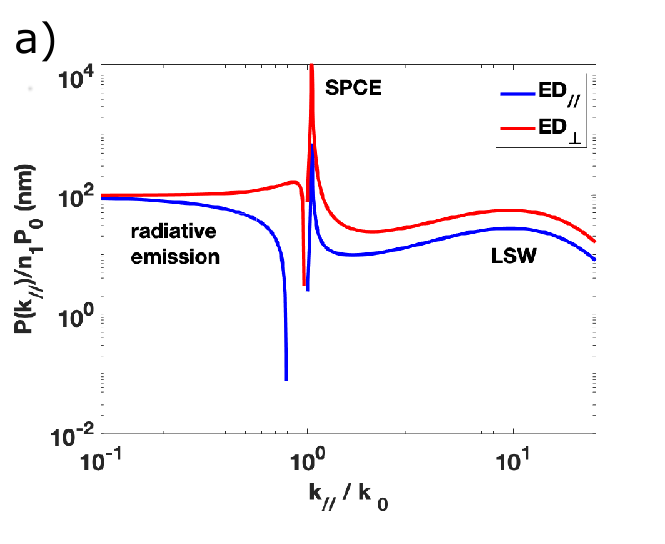}
\includegraphics[width=8cm]{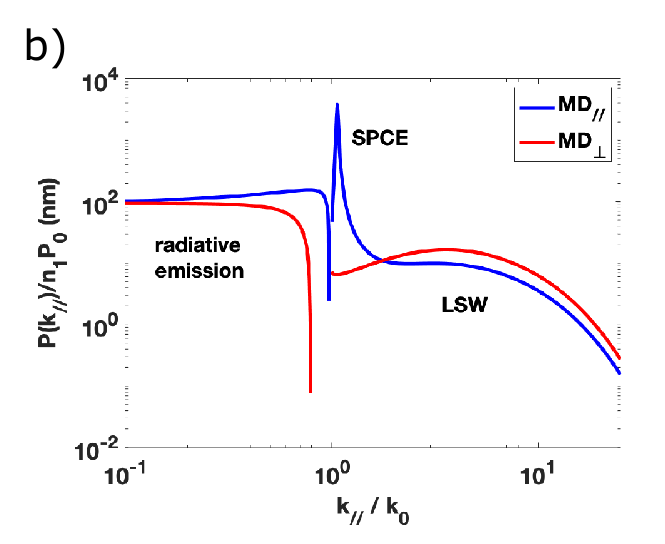}
	\caption{(a) Electric dipolar emission power spectrum $P (k_{//})$ above a gold mirror at the wavelength $\lambda=620$ nm. The different relaxation channels are indicated: radiative emission, surface plasmon coupled emission (SPCE) and lossy surface waves (LSW). ED parallel or perpendicular to the surface are considered. (b) Same as (a) for a magnetic dipole at the wavelength $\lambda=590$ nm. The dipole emitter is located 10 nm above the mirror. 
		\label{fig:ED-MDPower}}
\end{figure*}
Fluorescence lifetime of an emitter depends on the distance to the mirror, as demonstrated by Drexhage \cite{drexhage74,Barnes1998b}. ED and MD emissions present different behaviours as a function of the distance to the mirror \cite{Lukosz1979,Noginova:13,NorrisPRL:18}. We represent in Fig. \ref{fig:ED-MDPower} the different relaxation channels deduced from Sommerfeld expansion of the dipolar emission as a function of the in-plane wavector $k_{//}$ \cite{GCF-Barthes-Girard:2016}
\begin{eqnarray}
P=\int_0^\infty P(k_{//})dk_{//}
\end{eqnarray}
The power spectrum $P(k_{//})$ is presented in Fig. \ref{fig:ED-MDPower} for ED and MD emitters. Radiative emission corresponds to the range $0\le k_{//}\le k_0$. The Lorentzian peak at $k_{//}/k_0=n_{SPP}=1.05$ originates from surface plasmon coupled emission (SPCE) and contribution at high $k_{//}$ are associated to lossy surfaces waves (LSW) \cite{ford84}. We notably observe that SPCE occurs for all the cases except for a MD perpendicular to the surface. The relaxation channel for perpendicular MD is governed by the non radiative losses (LSW). Polarized ED/MD transitions of NaYF$_4$:Eu$^{3+}$ differently couple to surface plasmon polaritons (SPP) and would be of strong interest for SPP routing \cite{NorrisPRL:18}.  

\begin{figure*}[h!]
\includegraphics[width=8cm]{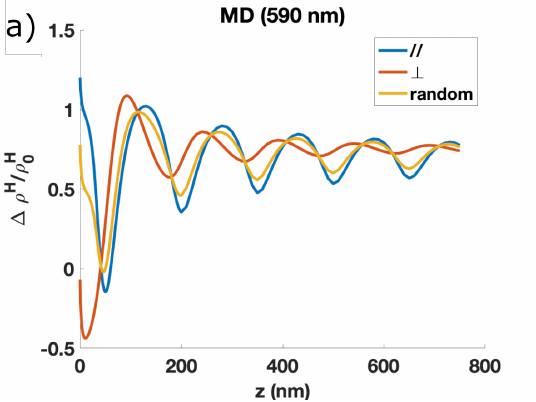}
\includegraphics[width=8cm]{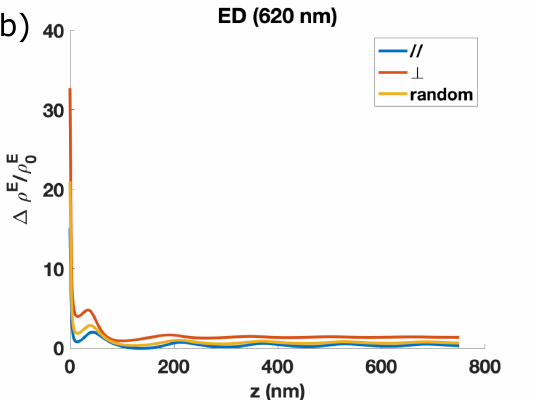}
\includegraphics[width=8cm]{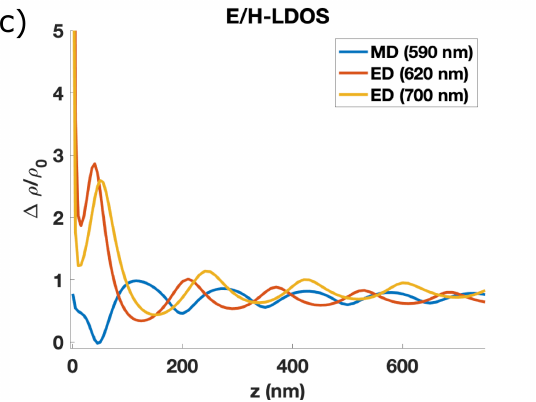}
	\caption{a,b) Variation of the normalized magnetic and electric LDOS at the wavelengths of interest.
	The contribution of the total (random) and partial LDOS either parallel or perpendicular to the surface are represented. c) Comparison of LDOS for the three optical transitions wavelengths: $^5D_0 \rightarrow ^7F_1$ (590 nm), $^5D_0 \rightarrow ^7F_2$ (620 nm) and $^5D_0 \rightarrow ^7F_4$ (700 nm). LDOS are averaged over the rod diameter (100 nm).  
		\label{fig:LDOS}}
\end{figure*}

Electric and magnetic LDOS correspond to the total ED or MD emission in the complex environment and are obtained by integrating the Sommerfeld expansion of Fig. \ref{fig:ED-MDPower}. Normalized LDOS are expressed as a function of electric and magnetic Green's tensor, respectively: $\widetilde{\Delta\rho}^{(E,H)}=6\pi/k_0^3 Im \left[ G^{(EE,HH)} \right]$. $G^{EE}$ refers to the electric Green's tensor above the mirror and its expression is given in ref. \cite{GCF-Barthes-Girard:2016}.  $G^{HH}$ refers to the magnetic Green's tensor above the mirror and is obtained from $G^{EE}$ with the exchange of the Fresnel coefficients $r_s \leftrightarrow r_p$. Note that we can safely neglect quantum effects (nonlocality, electronic spill-out, and Landau damping) on the metal optical response since their effects are not significative above few nanometers \cite{Corni:2003,Mortensen:20}. Normalized electric and magnetic LDOS, averaged over the rod height, are presented in Fig. \ref{fig:LDOS}. The variation of the magnetic LDOS is of the order of the free space LDOS whereas the electric LDOS is enhanced up to a factor of 30, as expected from stronger ED contribution to light-matter interaction in non magnetic materials.

The decay rate of the excited states $^5D_0$ is governed by ED ($^5$D$_0 \rightarrow$ $^7$F$_2$ and $^5$D$_0 \rightarrow ^7$F$_4$) and MD ($^5$D$_0 \rightarrow ^7$F$_1$) transitions. It is worth noticing that the emitter is a linear combination of incoherent electric and magnetic dipoles so that their decay rate contributions can be independently summed \cite{ZambranaPuyalto-Bonod:2015}. 
\begin{eqnarray}
\label{eq:decay0}
\Gamma_{D_0}&=&\Gamma_{^5D_0 \rightarrow ^7F_1}+\Gamma_{^5D_0 \rightarrow ^7F_2}+\Gamma_{^5D_0 \rightarrow ^7F_4} \\
\nonumber
&=&\Gamma_{0}+\Delta\Gamma_{^5D_0 \rightarrow ^7F_1}+\Delta\Gamma_{^5D_0 \rightarrow ^7F_2}+\Delta\Gamma_{^5D_0 \rightarrow ^7F_4} 
\end{eqnarray}
$\Gamma_{0}$ is the decay rate in free-space (including intrinsic non radiative rate). As explained above, we simplify the modelization merging the $^5$D$_0 \rightarrow$ $^7$F$_2$ and $^5$D$_0 \rightarrow ^7$F$_4$ contributions to a global electric contribution since they present similar ED dipole orientation. Therefore, decay rates modification induced by the optical environnement simplifies to   
\begin{eqnarray}
\label{eq:decay}
&&\Gamma_{D_0}=\Gamma_{0}+\Delta\Gamma_{MD}+\Delta\Gamma_{ED}\\
\nonumber
\\
\nonumber
&&\frac{\Delta\Gamma_{MD}}{\Gamma_{0}}=\sum_{\alpha=72^\circ,37^\circ}\eta_{\alpha}
\left\{ \left[\cos^2 \alpha+\frac{\sin^2\alpha }{2} \right]\widetilde{\Delta\rho}_{//}^H(d,d,\omega_{01}) \right. \\
\nonumber
&&\hspace{7cm} \left.+\frac{\sin^2\alpha }{2} \widetilde{\Delta\rho}_{\perp}^H(d,d,\omega_{01})\right\} \\
\nonumber
\\
\nonumber
&&\frac{\Delta\Gamma_{ED}}{\Gamma_{0}}=\chi_{ED}^2\sum_{\alpha=60^\circ,44^\circ}\eta_{\alpha}
\left\{ \left[\cos^2 \alpha+\frac{\sin^2\alpha }{2} \right]\widetilde{\Delta\rho}_{//}^E(d,d,\omega_{02}) \right. \\
\nonumber
&&\hspace{7cm} 
\left.+\frac{\sin^2\alpha }{2} \widetilde{\Delta\rho}_{\perp}^E(d,d,\omega_{02}) \right\} 
\end{eqnarray}
where $\chi_{ED}=3/(2+n_{rod}^2)$ is the local field correction inside the rod of optical index $n_{rod}=1.48$ \cite{Rabouw-Norris:16}. $\Delta\Gamma$ refers to the modification of the decay rate in front of the mirror 
and is expressed as a function of the transition quantum yields $\eta$ and partial LDOS.  We also present in Fig. \ref{fig:LDOS}c) the variation of the electric and magnetic LDOS at the emission wavelength of $^5$D$_0 \rightarrow$ $^7$F$_1$, $^5$D$_0 \rightarrow$ $^7$F$_2$ and $^5$D$_0 \rightarrow ^7$F$_4$ optical transitions. We observe that electric LDOS variations are similar in the very near-field of the mirror at 620 and 700 nm. This justifies to merge all ED presenting similar orientations.

\subsection{Drexhage's experiment}
In the following, we measure the lifetime of Europium doped nanorods as probe of the electromagnetic LDOS in a Drexhage's experiment. The confocal microscope was adapted to measure the lifetime of the photoluminescence of the Eu$^{3+}$-doped nanorods, as shown in Fig. \ref{fig:lifetimeSetup}. The europium ions were excited by a laser emitting $20 \mu s$ pulses at a wavelength of $392\pm 0.5$ nm, modulated at 10 Hz. The photoluminescence emission was collected through the same objective and directed towards an avalanche photodiode (APD).  Because of parasitic luminescence of impurities in SiO films, we first checked that we measured Eu$^{3+}$ emission recording a spectrum in CW regime before modulating the laser for decay time acquisition. The emitted signal was filtered (band pass
 filter range between 586.8 nm and 598.6 nm - Semrock FF01-600/14, Semrock FF02-586/15)
to acquire signal emitted 
from the the $^5D_0$ excited state only (avoiding notably the $^5D_1\rightarrow ^7F_3$ transitions at 580 nm).

\begin{figure*}[h!]
\begin{subfigure}{.4\textwidth}
\includegraphics[width=8cm]{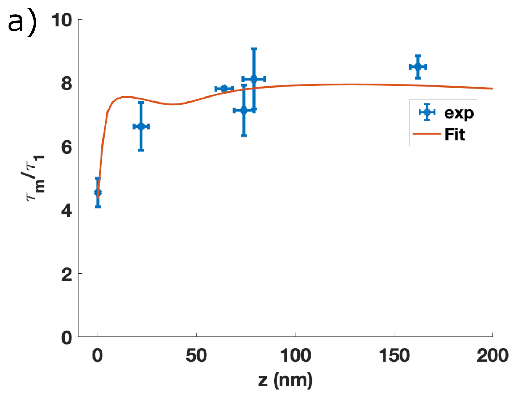}
\end{subfigure}
\hspace{.5cm}
\begin{subfigure}{.4\textwidth}
\includegraphics[width=8cm]{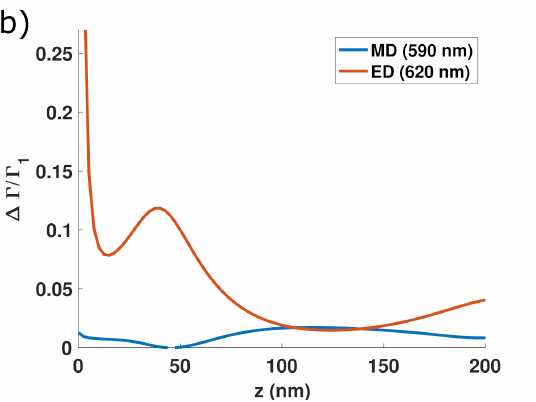}
\end{subfigure}
	\caption{a) $^5D_0$ fluorescence lifetime as a function of the distance to the gold mirror. Solid line is a fit with Eq. \ref{eq:decay}. b) MD and ED contributions to the total decay rate. Fitting parameters are $\tau_0=1/\Gamma_0=8.2$ ms and $A=0.12$.	
\label{fig:lifetime5D0}}
\end{figure*}

Quantum yield is proportionnal to the oscillator strength \cite{Hovhannesyan-Lepers:21} so that we can rewrite the above expression (Eq. \ref{eq:decay}) with $\eta_\alpha =A f_\alpha$ where $A$ is a proportionality factor (that depends on the materials). The only unknown parameters are the free space decay rate $\Gamma_0$ and the factor $A$. Fig. \ref{fig:lifetime5D0} shows the fluorescence lifetime measured as a function of the distance to the mirror and the corresponding fitting curve. The MD and ED contributions to the decay rate (inverse of lifetime) are also presented in fig. \ref{fig:lifetime5D0}b).  We clearly observe that the main contribution to the decay rate originates from the ED transitions at short distances, where the variations are large. It is worth mentioning that the large LDOS observed in the very near-field of the gold mirror is associated to SPCE and LSW, that are not accessible from radiative LDOS, estimated by recording the fluorescence intensity. In addition, at short distances, the variations due to $^5D_0 \rightarrow ^7F_2$ (ED$_1$) and $^5D_0 \rightarrow ^7F_4$ (ED$_2$) transitions are similar as shown in Fig. \ref{fig:LDOS}c, and considering a unique ED transition may satisfactorily reproduces the evolution of the decay rate in the near-field of the  mirror \cite{Kunz-Lukosz:80}.  From the fitting parameters, one can also deduce the intrinsic non radiative rate $\Gamma_0^{nr}$ as well as the free space radiative decay rates associated to each transition, see table \ref{tab:fitDecay2}. We found a quantum yield of 20\%.

\begin{table} [!ht]
\caption{Parameters deduced from the fitting parameters. $\Gamma_0^{nr}$: intrinsic non radiative decay rate;  
$\Gamma_{MD}^{rad}$ ($\Gamma_{ED}^{rad}$): radiative electric (magnetic) decay rate; QY: quantum yield.} 
\begin{tabular}{cccc}
$\Gamma_0^{nr}$	& $\Gamma_{MD}^{rad}$& $\Gamma_{ED}^{rad}$& QY	\\ 
\hline
103 s$^{-1}$ &7 s$^{-1}$& 14 s$^{-1}$ & 20\% \\
\end{tabular}
\label{tab:fitDecay2}
\end{table}

\section{Conclusion}
We investigated in detail optical emission of NaYF$_4$:Eu$^{3+}$ nanorods in complex environments. In a first step, measuring emitted light intensity permits to determine the intrinsic optical properties of europium ions emitters in single crystalline NaYF$_4$ nanorod host matrix;  namely the dipole moment nature, orientation and relative oscillator strengths of the $^5D_0 \rightarrow ^7F_J$ optical transitions. Crystal field splitting of the $^7$F$_J$ level leads to $2J+1$ Stark sublevels spectrally separated. We fully resolved the three Stark sublevels for the MD transitions ($J=1$) and some of them for the two ED transitions:  two of the five Stark sublevels for  $J=2$ and   five of the nine Starks sublevels for $J=4$.  
Europium doped nanorods present well defined ED/MD transitions of strong interest as vectorial nanoprobes to investigate the modification of electric and magnetic LDOS in complex environnements. Since the dipolar  emitters present fixed orientation with respect to the rod long axis, they can be used as vectorial nanoprobes of the electromagnetic LDOS, as demonstrated in the second part of this work. We determined that LDOS in front of a gold mirror is mainly of electric nature. We also estimated the radiative and non radiative decay rates for both ED and MD transitions and a global quantum yield of 20\%. This demonstrates rare-earth doped single crystaline NaYF$_4$ as versatile nanoprobes of their electromagnetic environment and reciprocically that their electromagnetic environment can be used to probe their optical intrinsic properties. Further engineering of these nanoprobes would be helpfull to control the ED/MD polarization and mapping specific partial LDOS. Such nanoprobes constitute a powerful tool to characterize thermal emission and electromagnetic properties in optoelectronics devices. Deep understanding of the electric and magnetic contribution would notably help in understanding and controlling near-field radiative transfer towards low power components. 
\ack
 We gratefully acknowledge Fr\'ed\'eric Herbst from the technological platform ARCEN Carnot for SEM imaging of single nanorods.
 
Platform ARCEN Carnot is financed by the R\'egion de Bourgogne Franche-Comt\'e and the D\'el\'egation R\'egionale \`a la Recherche et \`a la Technologie (DRRT). This work has benefited from the facilities of the SMARTLIGHT platform in Bourgogne Franche-Comté (EQUIPEX+ contract "ANR-21-ESRE-0040") and is supported by the French Investissements d'Avenir program EUR-EIPHI (17-EURE-0002), French National Research Agency (ANR) project SpecTra (ANR-16-CE24-0014) and the European Regional Development Fund  FEDER-FSE 2014-2020, Bourgogne-Franche-Comté.



\section*{References}
\begin{thebibliography}{10}

\bibitem{joulain05}
K.~Joulain, J.P. Mulet, F.~Marquier, R.~Carminati, and J.J. Greffet.
\newblock Surface electromagnetic waves thermally excited : radiative heat
  transfer, coherence properties and casimir forces revisited in the near
  field.
\newblock {\em Surf. Sci. Rep.}, 57:59--112, 2005.

\bibitem{Moskalensky-Yurkin:21}
A.~E. Moskalensky and M.~A. Yurkin.
\newblock A point electric dipole: From basic optical properties to the
  fluctuation-dissipation theorem.
\newblock {\em Reviews in Physics}, 6:100047, 2021.

\bibitem{Barnes1998b}
W.L. Barnes.
\newblock Fluorescence near interfaces: the role of photonic mode density.
\newblock {\em Journal of Modern Optics}, 45:661--699, 1998.

\bibitem{GCF-Barthes-Girard:2016}
G.~{Colas des Francs}, J.~Barthes, A.~Bouhelier, J-C. Weeber,
  A.~Dereux, A.~Cuche, and C.~Girard.
\newblock Plasmonic purcell factor and coupling efficiency to surface plasmons.
  implications for addressing and controlling optical nanosources.
\newblock {\em Journal of Optics}, 18:094005, 2016.

\bibitem{Wojszvzyk:19}
L.~Wojszvzyk, H.~Monin, and J.-J Greffet.
\newblock Light emission by a thermalized ensemble of emitters coupled to a
  resonant structure.
\newblock {\em Advanced Optical Materials}, page 1801697, 2019.

\bibitem{Cuevas-GarciaVidal:18}
J.~C. Cuevas and F.~J. Garc\'a-Vidal.
\newblock Radiative heat transfer.
\newblock {\em ACS photonics}, 5:3896--3915, 2018.

\bibitem{Lodahl:2015}
P.~Lodahl, S.~Mahmoodian, and S.~Stobbe.
\newblock Interfacing single photons and single quantum dots with photonic
  nanostructures.
\newblock {\em Review of Modern Physics}, 87:347--400, 2015.

\bibitem{Dereux-Devaux-Weeber-Goudonnet-Girard:2001}
A.~Dereux, E.~Devaux, J.~C. Weeber, J.~P. Goudonnet, and C.~Girard.
\newblock Direct interpretation of near--field optical images.
\newblock {\em Journal of Microscopy}, 202:320--331, 2001.

\bibitem{NanotechDereuxGCF:2003}
A.~Dereux, C.~Girard, C.~Chicanne, {G. {Colas des Francs}}, T.~David,
  Y.~Lacroute, and J.-C. Weeber.
\newblock Subwavelength mapping of surface photonic states.
\newblock {\em Nanotechnology}, 14:935, 2003.

\bibitem{deWilde:06}
Y.~de~Wilde, F.~Formanek, R.~Carminati, B.~Gralak, P.A. Lemoine, K.~Joulain,
  J.P. Mulet, Y.~Chen, and J.J. Greffet.
\newblock Thermal radiation scanning tunnelling microscopy.
\newblock {\em Nature}, 444:740, 2006.

\bibitem{HoogenboomNanoLet:2009}
Jacob~P. Hoogenboom, Gabriel {Sanchez-Mosteiro}, G.~{Colas des
  Francs}, Dominique Heinis, Guillaume Legay, Alain Dereux, and Niek~F. van
  Hulst.
\newblock The single molecule probe: nanoscale vectorial mapping of photonic
  mode density in a metal nanocavity.
\newblock {\em Nano Letters}, 9:1189--1195, 2009.

\bibitem{Agio:2012}
Mario Agio.
\newblock Optical antennas as nanoscale resonators.
\newblock {\em Nanoscale}, 4:692--706, 2012.

\bibitem{Kasperczyk:15}
M.~Kasperczyk, S.~Person, D.~Ananias, L.~D. Carlos, and L.~Novotny.
\newblock Excitation of magnetic dipole transitions at optical frequencies.
\newblock {\em Physical Review Letters}, 114:163903, 2015.

\bibitem{Pham-Drezet:16}
A.~Pham, M.~Berthel, Q.~Jiang, J.~Bellessa, S.~Huant, C.~Genet, and A.~Drezet.
\newblock Chiral optical local density of states in a spiral plasmonic cavity.
\newblock {\em Physical Review A}, 94:053850, 2016.

\bibitem{Karaveli-Zia:2011}
S.~Karaveli and R.~Zia.
\newblock Spectral tuning by selective enhancement of electric and magnetic
  dipole emission.
\newblock {\em Physical Review Letters}, 106:193004, 2011.

\bibitem{Aigouy:14}
L.~Aigouy, A.~Caz\'e, P.~Gredin, M.~Mortier, and R.~Carminati.
\newblock Mapping and quantifying electric and magnetic dipole luminescence at
  the nanoscale.
\newblock {\em Physical Review Letters}, 113(076101), 2014.

\bibitem{Rabouw-Norris:16}
F.~T. Rabouw, P.~T. Prins, and D.~J. Norris.
\newblock Europium-doped {NaYF$_4$} nanocrystals as probes for the electric and
  magnetic local density of optical states throughout the visible spectral
  range.
\newblock {\em Nano Letters}, 16:7254--7260, 2016.

\bibitem{Li-Zia:18}
D.~Li, S.~Karaveli, S.~Cueff, W.~Li, and R.~Zia.
\newblock Probing the combined electromagnetic local density of optical states
  with quantum emitters supporting strong electric and magnetic transitions.
\newblock {\em Physical Review Letters}, 121:227403, 2018.

\bibitem{Wiecha-Cuche:19}
P.~Wiecha, C.~Majorel, C.~Girard, A.~Arbouet, B.~Masenelli, O.~Boisron,
  A.~Lecestre, G.~Larrieu, V.~Paillard, and A.~Cuche.
\newblock Enhancement of electric and magnetic dipole transition of
  rare-earth-doped thin films tailored by high-index dielectric nanostructures.
\newblock {\em Applied Optics}, 58:1682--1690, 2019.

\bibitem{Bidault-Mivelle-Bonod:19}
S.~Bidault, M.~Mivelle, and N.~Bonod.
\newblock Dielectric nanoantennas to manipulate solid-state light emission.
\newblock {\em Journal of Applied Physics}, 126(094104), 2019.

\bibitem{Suyver-Gudel:06}
J.~F. Suyver, J.~Grimm, M.~van Veen, D.~Biner, K.~Kr\"amer, and H.-U. G\"udel.
\newblock Upconversion spectroscopy and properties of nayf$_4$ doped with
  er$^{3+}$, tm${^3+}$ and/or yb$^{3+}$.
\newblock {\em Journal of luminescence}, 117:1--12, 2006.

\bibitem{Chacon-GCF:2020}
R.~Chacon, A.~Leray, J.~Kim, K.~Lahlil, S.~Mathew, A.~Bouhelier, J.-W. Kim,
  T.~Gacoin, and {G. {Colas des Francs}}.
\newblock Measuring the magnetic dipole transition of single nanorods by
  spectroscopy and fourier microscopy.
\newblock {\em Physical Review Applied}, 14:054010, 2020.

\bibitem{Kim-Gacoin:21}
J.~Kim, R.~Chacon, Z.~Wang, E.~Larquet, K.~Lahlil, A.~Leray, {G.
  {Colas des Francs}}, J.~Kim, and T.~Gacoin.
\newblock Measuring 3D orientation of nanocrystals via polarized luminescence
  of rare-earth dopants.
\newblock {\em Nature Communications}, 12:1943, 2021.

\bibitem{Kumar-Fick:20}
A.~Kumar, A.~Asadollahbaik, J.~Kim, K.~Lahlil, S.~Thiele, K.~Weber,
  J.~Drozella, F.~Sterl, N.S. Chormaic, A.M. Herkommer, J.~Kim, T.~Gacoin,
  H.~Giessen, and J.~Fick.
\newblock Optical trapping and orientation-resolved spectroscopy of europium-doped nanorods
\newblock {\em Journal of Physics Photonics}, 2:025007, 2020.

\bibitem{Sayre-Freed:1956}
E.~V. Sayre and S.~Freed.
\newblock Spectra and quantum states of the europic ion in crystals.
  {II.F}luorescence and absorption spectra of single crystals of europic
  ethylsulfate nonahydrate.
\newblock {\em The Journal of Chemical Physics}, 24:1213--1219, 1956.

\bibitem{Kim-Gacoin:17}
J.~Kim, S.~Michelin, M.~Hilbers, L.~Martinelli, E.~Chaudan, G.~Amselem,
  E.~Fradet, J.-P. Boilot, A.~M. Brouwer, C.N. Baroud, J.~Peretti, and
  T.~Gacoin.
\newblock Monitoring the orientation of rare-earth-doped nanorods for flow
  shear tomography.
\newblock {\em Nature Nanotechnology}, 12:914--919, 2017.

\bibitem{Sick-Hecht-Wild-Novotny:2001}
B.~Sick, B.~Hecht, U.~P. Wild, and L.~Novotny.
\newblock Probing confined fields with single molecules and vice versa.
\newblock {\em Journal of Microscopy}, 202:365--373, 2001.

\bibitem{Buchler-Kalkbrenner-Hettich-Sandoghdar:2005}
B.~C. Buchler, T.~Kalkbrenner, C.~Hettich, and V.~Sandoghdar.
\newblock Measuring the quantum efficiency of the optical emission of single
  radiating dipoles using a scanning mirror.
\newblock {\em Physical Review Letters}, 95:63004, 2005.

\bibitem{Wang-Liu:2010}
F.~Wang, Y.~Han, C.~S. Lim, Y.~Lu, J.~Wang, J.~Xu, H.~Chen, C.~Zhang, M.~Hong,
  and X.~Liu.
\newblock Simultaneous phaseand size control of upconversion nanocrystals
  through lanthanide doping.
\newblock {\em Nature}, 463:1061--1065, 2010.

\bibitem{Binnemans:96}
K.~Binnemans:96.
\newblock Interpretation of Europium (III) spectra. 
\newblock {\em Coordination Chemistry Reviews}, 295:1--45, 1996.

\bibitem{Tu-Chen:13}
D.~Tu, Y.~Liu, H.~Zhu, R.~Li, L.~Liu, and X.~Chen.
\newblock Breakdown of crystallographic site symmetry in lanthanide-doped
  NaYF$_4$ crystals.
\newblock {\em Angewandte Communications}, 52:1128--1133, 2013.

\bibitem{Hass-Salzberg:54}
G.~Hass and C.~D. Salzberg.
\newblock Optical properties of silicon monoxide in the wavelength region from
  0.24 to 14.0 microns.
\newblock {\em J. Opt. Soc. Am.}, 44(181-183), 1954.

\bibitem{Johnson-Christy:1972}
P.B. Johnson and R.W. Christy.
\newblock Optical constants of the noble metals.
\newblock {\em Physical Review B}, 6:4370--4379, 1972.

\bibitem{Lukosz1979}
W.~Lukosz.
\newblock Light emission by magnetic and electric dipoles close to a plane
  dielectric interface. iii radiation patterns of dipoles with arbitrary
  orientation.
\newblock {\em J. Opt. Soc. Am.}, 69:1495--1503, 1979.

\bibitem{drexhage74}
K.~H. Drexhage.
\newblock Interaction of light with monomolecular dye layers.
\newblock In E.~Wolf, editor, {\em Progress in Optics}, volume~12, pages
  161--232. North Holland, Amsterdam, 1974.

\bibitem{Noginova:13}
N.~Noginova, R.~Hussain, M.~A. Noginov, J.~Vella, and A.~Urbas.
\newblock Modification of electric and magnetic dipole emission in anisotropic
  plasmonic systems.
\newblock {\em Optics Express}, 21:23087--23096, 2013.

\bibitem{NorrisPRL:18}
R.~Brechb\"uhler, F.~T. Rabouw, P.~Rohner, B.~le~Feber, D.~Poulikakos, and
  D.~J. Norris.
\newblock Two-dimensional drexhage experiment for electric- and magnetic-dipole
  sources on plasmonic interfaces.
\newblock {\em Physical Review Letters}, 121:113601, 2018.

\bibitem{ford84}
G.~W. Ford and W.~H. Weber.
\newblock Electromagnetic interactions of molecules with metal surfaces.
\newblock {\em Physics Reports}, 113:195--287, 1984.

\bibitem{Corni:2003}
S. Corni and J. Tomasi
\newblock Lifetimes of electronic excited states of a molecule close to a metal surface
\newblock {\em J. Chem. Phys.} ,118, 6481 2003.

\bibitem{Mortensen:20}
P.~Gonçalves and T.~Christensen and N.~Rivera and A.-P. Jauho and N. Asger Mortensen and M. Soljacic.
\newblock Plasmon–emitter interactions at the nanoscale.
\newblock {\em Nature Communications}, 11: 366, 2020.

\bibitem{ZambranaPuyalto-Bonod:2015}
X.~Zambrana-Puyalto and N.~Bonod.
\newblock Purcell factor of spherical mie resonators.
\newblock {\em Physical Review B}, 91:195422, 2015.

\bibitem{Hovhannesyan-Lepers:21}
G. Hovhannesyan, V. Boudon, and M. Lepers.
\newblock Transition intensities of trivalent lanthanide ions in
  solids: Extending the Judd-Ofelt theory.
\newblock {\em Journal of Luminescence},  241:118456, 2021.

\bibitem{Kunz-Lukosz:80}
R.E. Kunz and W.~Lukosz.
\newblock Changes in fluorescence lifetimes induced by variable optical
  environments.
\newblock {\em Physical Review B}, 21:4814--4828, 1980.

\end{thebibliography}
\end{document}